\documentclass[twocolumn]{revtex4}
\usepackage{amsmath,amssymb}
\usepackage{natbib}
\usepackage{url}
\usepackage{graphicx}
\usepackage{psfrag}
\usepackage{color}
\usepackage{epstopdf}

\usepackage{mathrsfs}
\DeclareSymbolFontAlphabet{\mathrsfs}{rsfs}
\DeclareMathAlphabet{\mathcal}{OMS}{cmsy}{m}{n}

\newcommand{\scri}{\mathrsfs{I}}
\newcommand{\be}{\begin{equation}}
\newcommand{\ee}{\end{equation}}

\def\tg{{\tilde{g}}}

\def\rhoo{\rho_{\mathrm{out}}}
\def\rhoi{\rho_{\mathrm{in}}}

\begin{document}
\title{Hyperboloidal evolution of test fields in three spatial dimensions}

\author{An{\i}l Zengino\u{g}lu$^{1}$} 
\author{Lawrence E. Kidder$^2$} 

\affiliation{$^1$ M.~Smoluchowski Institute of Physics, Jagiellonian University, Krak\'ow, Poland}

\affiliation{$^2$ Center for Radiophysics and Space Research, Cornell University, Ithaca, New York, 14853}

\begin{abstract}
We present the numerical implementation of a clean solution to the outer boundary and radiation extraction problems within the 3+1 formalism for hyperbolic partial differential equations on a given background. Our approach is based on compactification at null infinity in hyperboloidal scri fixing coordinates. We report numerical tests for the particular example of a scalar wave equation on Minkowski and Schwarzschild backgrounds. We address issues related to the implementation of the hyperboloidal approach for the Einstein equations, such as nonlinear source functions, matching, and evaluation of formally singular terms at null infinity.
\end{abstract}

\pacs{}
\maketitle

\section{Introduction}
The most common method in the numerical construction of solutions to hyperbolic partial differential equations (PDEs) 
is the 3+1 formalism. In this approach the spacetime is foliated by a family of spacelike hypersurfaces that are level sets of a time function. Typically, the spatially unbounded domain is truncated by an artificial timelike outer boundary. The solution is then calculated numerically on a finite, spatially compact domain with given initial and boundary data. In this context, we present a successful numerical implementation of a clean solution to difficulties related to (i) the construction and imposition of outer boundary conditions, and (ii) the extraction of outgoing radiation for test fields.  

The first difficulty arises due to the artificial nature of the timelike boundary, which is not part of the physical problem. Ideally, boundary conditions ensure transparency of the domain boundary; however, spurious reflections occur from the outer boundary even for the simple case of the flat wave equation in a three-dimensional ball \cite{Sarbach:2007hd}. Boundary conditions that are constructed to minimize spurious reflections are called nonreflecting, absorbing or radiation boundary conditions \cite{Givoli91, Hagstrom99, Hagstrom07}. They involve approximations to ideal conditions and substantial work goes into error controlling. In particular, it is difficult to construct, improve, or implement accurate absorbing boundary conditions when the background curvature does not vanish or when non-linear terms appear in the equations. One needs to account for backscatter off of curvature or self-interaction of the field near the boundary. This is relevant because a bad choice of boundary data can reduce the accuracy and even destroy certain features of the solution. Especially in long time simulations, the outer boundary can be a significant limiting factor for the accuracy of numerical computations.

The second difficulty is related to the notion of radiation. There are three scales of interest in radiative systems: the scale of the source, the wavelength of the emitted radiation, and the location of the observer. Radiation is defined asymptotically in the far field zone where the observer is located.  To study radiation accurately, one needs to evolve large portions of spacetime while keeping high resolution near the source. This is inefficient with Cauchy foliations because the resolution in the outer domain is unnecessarily high even with mesh refinement techniques, and the time lag between the source and the observer increases proportionally to their distance.

A clean solution to these difficulties is to avoid the artificial truncation of the computational domain by compactification. It is well known that compactification at spatial infinity is not compatible with hyperbolic PDEs \cite{Grandclement:2009}. It leads to a loss of resolution due to blueshift in the wavenumber of outgoing waves. Compactification needs to be performed with respect to the outgoing characteristic direction of the hyperbolic PDE. In the 3+1 formalism a suitable time transformation needs to be applied  to standard Cauchy surfaces before compactification so that the time surfaces approach null infinity in the asymptotic domain. Such time surfaces are called hyperboloidal because their causal behavior resembles that of standard hyperboloids in Minkowski spacetime \cite{Friedrich83a}.

Until now, numerical studies of the hyperboloidal initial value problem for test fields have been restricted to one or two dimensional codes \cite{Fodor:2003yg, Fodor:2006ue, Bizon:2008zd, Hennig:2008af, Zenginoglu:2008uc, Zenginoglu:2008wc, Zenginoglu:2009hd, Zenginoglu:2009ey} with the exception of \cite{vanMeter:2006mv}. The hyperboloidal approach has been shown to be accurate and efficient in lower dimensional studies, but in \cite{vanMeter:2006mv} the authors report large numerical errors in the simulation of Maxwell fields across future null infinity in flat spacetime with a 3D code based on Cartesian coordinates. They deal with this problem by adding an artificial cosmological term to the background and by artificially tilting the light cones beyond null infinity to make grid boundaries spacelike. They suggest, however, that the errors may not be present in a numerical code that supports a spherical grid topology. We confirm this expectation in scri fixing coordinates with the Spectral Einstein Code {\texttt SpEC} \cite{SpECWebsite}.

A strong motivation for the hyperboloidal approach is to improve the accuracy and efficiency of numerical relativistic calculations of binary black holes where the outer boundary and the radiation extraction problems have distinctive features \cite{Buchman:2007pj, Rinne:2008vn, Friedrich:2009tq, Seiler:2008hm, Nunez:2009wn, Boyle:2009vi, Pazos:2006kz, Hannam:2007ik}. There is an ongoing effort to solve a hyperboloidal initial value problem for various versions of the Einstein equations \cite{Friedrich81b, Husa01, Frauendiener04, Friedrich04,Zenginoglu:2008pw, Moncrief:2008ie, Rinne:2009}. 
Our experiments have been chosen to address issues that appear in these calculations.

\section{Hyperboloidal scri fixing compactification}
\label{sec:2}
It is favorable for numerical purposes to map null infinity (scri) to a fixed coordinate sphere \cite{Frauendiener98b, Husa:2005ns}. Hyperboloidal scri fixing coordinates for asymptotically flat spacetimes have been presented in \cite{Zenginoglu:2007jw}. Starting from the representation of a background spacetime based on Cauchy surfaces, hyperboloidal scri fixing compactification consists of: (i) a time transformation, (ii) a radial compactification, and (iii) a rescaling. 

The first class of hyperboloidal surfaces studied are constant mean curvature (CMC) surfaces \cite{Eardley79,York79}. Brill, Cavallo and Isenberg constructed spherically symmetric CMC surfaces with nonvanishing mean extrinsic extrinsic curvature in Schwarzschild spacetime \cite{Brill80}. Recently, Malec and O'Murchadha analyzed such surfaces in great detail \cite{Malec:2003dq,Malec:2009hg}. CMC slicings serve as the starting point for the analysis of hyperboloidal slicings of the Schwarzschild spacetime defined by members of the Bona-Mass\'o family of slicing conditions \cite{Ohme:2009gn}. Beyond Schwarzschild spacetime, the CMC condition plays a central role in the construction of hyperboloidal initial data for Einstein equations presented by Buchman, Bardeen and Pfeiffer \cite{Buchman:2009ew}. 

We employ CMC foliations in most of the numerical experiments. However, the CMC condition may be regarded as restrictive due to its local nature, whereas the hyperboloidal condition for spacelike surfaces is only asymptotic. Therefore we discuss the steps in the construction of hyperboloidal scri fixing coordinates in a general manner for a given static background before presenting the explicit results.
\subsection{General construction on a static background}
The metric on a static background spacetime can be written as \be \label{eq:bgr}
\tilde{g}=\tilde{g}_{tt}dt^2+\tilde{g}_{rr}dr^2 + r^2\,d\sigma^2, \ee
where $d\sigma^2$ is the standard metric on the unit sphere. We introduce a new time function
\be \label{time_tr} \tau = t - h(r). \ee
Here, $h(r)$ is called the height function. This  transformation implies that the coordinates of the
foliation are adapted to time symmetry in the sense that the timelike
Killing vector is left invariant. We choose the height function so that level sets of $\tau$ asymptotically approach future null infinity instead of spatial infinity (see for example \eqref{eq:height_mink} and \eqref{eq:height_ss}). 

The outgoing spatial direction is compactified by the radial transformation
\be \label{rad_tr} r = \frac{\rho}{\Omega(\rho)}. \ee
We choose the function $\Omega$ such that it vanishes at a finite coordinate value with respect to $\rho$ (see for example (\ref{eq:rho_mink}) or (\ref{ss_conf})). This implies that $\Omega$ vanishes at infinity with respect to $r$. The zero set of $\Omega$ corresponds to future null infinity denoted by $\scri^+$ (scri). 

The above compactification leads to a singular metric at the finite coordinate distance that corresponds to scri. This  singularity can be rescaled away with a suitable conformal factor if the spacetime is asymptotically simple \cite{Penrose63}. We introduce the conformal metric
\be\label{conf_tr} g = \Omega^2 \tilde{g}. \ee
The metric $g$ is not a solution to the vacuum Einstein equations due to the transformation behavior of the connection under a conformal rescaling of the metric \cite{Wald84}. 

The transformations (\ref{time_tr}), (\ref{rad_tr}), and (\ref{conf_tr}) lead to the conformal metric
\be\label{eq:stat_conf} g = \Omega^2 \tilde{g}_{tt}\,d\tau^2+ 
2\tilde{g}_{tt} H L\, d\tau d\rho +
\frac{\tilde{g}_{rr}+\tilde{g}_{tt}H^2}{\Omega^2}L^2\,d\rho^2+\rho^2d\sigma^2. \ee
We defined $H:=(dh/dr) (\rho)$ and
$L:=\Omega-\rho\Omega'$. A prime denotes derivative with respect to $\rho$. Not every choice of the height function derivative $H$ will result in a spacelike foliation. The height function derivative must satisfy 
\be\label{eq:sp_cond} H^2 < \frac{ \tilde{g}_{rr}}{-\tilde{g}_{tt}} \ee
The formal singularity of the conformal metric \eqref{eq:stat_conf} can be removed by asymptotic simplicity and by suitable choices of the involved functions as demonstrated explicitly for Minkowski and Schwarzschild spacetimes in the following subsections. 

We will often refer to the 3+1 decomposition of the conformal metric with respect to the time direction. Introducing the lapse $\alpha$, the shift $\beta$, and the spatial metric function $\gamma$, we write
\be\label{gen_decomp} g = \left(-\alpha^2 + \gamma^2 \beta^2\right)\,d\tau^2 + 2 \gamma^2\beta\, d\tau\,d\rho + \gamma^2 \,d\rho^2 + \rho^2 \,d\sigma^2. \ee

\subsection{Minkowski spacetime}\label{sec:mink}
The Minkowski metric $\tilde{\eta}$ in the standard spherical coordinate system reads
\be\label{eq:st_mink} \tilde{\eta} = -dt^2+dr^2+r^2\,d\sigma^2.\ee
We set the height function such that level sets of the new time coordinate $\tau$ are CMC surfaces:
\be \label{eq:height_mink}
 h(r) = \sqrt{\frac{a^2}{S^2}+r^2}, \quad H(r) = \frac{r}{\sqrt{\frac{a^2}{S^2}+r^2}}. \ee Here, $a$ and $S$ are free parameters. The mean extrinsic curvature of level sets of $\tau$ reads $K=3S/a$. Surfaces with positive $K$ approach future null infinity in our convention. The radial transformation is determined by the choice of the conformal factor. We set
  \be \label{eq:rho_mink} \Omega = \frac{S^2-\rho^2}{2a} \quad \Rightarrow \quad L =\frac{S^2+\rho^2}{2a} . \ee The set $\{\rho=S\}$ corresponds to future null infinity and $\{\rho=0\}$ corresponds to the origin. The conformal factor is a function of $\rho^2$ in Minkowski spacetime because the coordinate $\rho$ is not regular at the origin. From (\ref{eq:stat_conf}) we get for the conformal Minkowski metric
\be\label{eq:con_mink} \eta =
\Omega^2 \tilde{\eta} = -\Omega^2\,d\tau^2 - 2\rho
\frac{S}{a}\,d\tau\,d\rho+ d\rho^2+\rho^2\,d\sigma^2. \ee Lapse, shift
and the spatial metric function for the above metric are
\be\label{mink_decomp} \alpha = \frac{S^2+\rho^2}{2a},\qquad \beta = -\rho\frac{S}{a},\qquad \gamma = 1.\ee The Ricci scalar of the conformal metric reads
\[ R = \frac{12(S^2-\rho^2)(\rho^2+3S^2)}{(\rho^2+S^2)^3}. \] 

\subsection{Schwarzschild spacetime}\label{sec:ss}
The Schwarzschild metric $\tilde{g}$ with mass $m$ in the standard spherical coordinate system reads
\[ \tilde{g} = - F \, dt^2 + \frac{1}{F}\, dr^2 + r^2 \, d\sigma^2, \quad \textrm{with} \quad   F = 1-\frac{2m}{r}. \]
\subsubsection{CMC foliation}
The height function for a CMC foliation of Schwarzschild spacetime can not be written in explicit form. Its derivative is given by \cite{Brill80, Malec:2003dq,Malec:2009hg}
\be\label{eq:height_ss}
H=\frac{J} {F \sqrt{J^2+F}},  \qquad J:= \frac{K r}{3}-\frac{C}{r^2}.\ee
Here, $K$ is the mean
extrinsic curvature and $C$ is an integration constant. For numerical applications these parameters should satisfy $K>0$ and $C>8m^3 K/3$, which makes sure that the surfaces approach future null infinity and cross the future event horizon. 

A convenient choice for the conformal factor is
\be \label{ss_conf} \Omega=1-\frac{\rho}{S}\quad \Rightarrow \quad L=1.\ee
With $\bar{J}:=\Omega J$, we get for the conformal Schwarzschild metric 
\begin{align}  \label{eq:con_ss} g &= -\Omega^2F\,d\tau^2 - 2 \frac{\bar{J}}{\sqrt{\bar{J}^2+F\Omega^2}} \,d\tau\,d\rho  \nonumber \\
&+ \frac{1}{\bar{J}^2 + F \Omega^2}\,d\rho^2+\rho^2\,d\sigma^2. \end{align}
The functions from the 3+1 decomposition (\ref{gen_decomp}) read 
\be\label{ss_decomp} \alpha = \sqrt{\bar{J}^2+F\Omega^2},\quad \beta = 
-\bar{J} \alpha\,\qquad \gamma = \alpha^{-1} . \ee
The Ricci scalar reads
\[  R = \frac{12 \Omega (m(2\rho-S)+ \rho S)}{\rho^2 S^3}.  \] 

\subsubsection{Matching}\label{sec:match}
The success of numerical calculations in computing the inspiral of binary black hole systems suggests that methods developed for the treatment of the asymptotic domain should be independent of the interior evolution. Also, many of the sophisticated techniques dealing with discontinuities and shocks related to the presence of matter near the black holes use special techniques in specific coordinate systems. It is desirable to leave these methods untouched in an effort to solve problems related to the asymptotic domain.

An essential advantage of hyperboloidal foliations in this context is their flexibility. The only condition for a hyperboloidal foliation in compact domains is that it is spacelike. One can keep the numerical calculation near the sources untouched and change the foliation in the exterior domain to approach null infinity. In this setting, hyperboloidal coordinates are used only as a means to avoid boundary conditions and to have access to the idealized radiation signal \cite{Zenginoglu:2007jw, Zenginoglu:2009hd, Zenginoglu:2009ey}. 

We describe the matching of an ingoing Eddington-Finkelstein foliation near a Schwarzschild black hole to hyperboloidal coordinates in an exterior domain. The Schwarzschild spacetime with an arbitrary $H$ and $\Omega$ can be written as
\begin{align} \label{eq:ss_match} g =-\Omega^2F\, d\tau^2
- 2L\left(H-\frac{2m\Omega}{\rho}(1+H)\right) d\tau\,d\rho \nonumber \\
+ \frac{L^2}{\Omega^2} \left(1-H^2+\frac{2m\Omega}{\rho}(1+H)^2\right)\,d\rho^2 
+ \rho^2 \, d\sigma^2. \end{align}
The functions of the 3+1 decomposition are
\begin{align}\label{eq:match_3p1} \alpha = \frac{\Omega}{\sqrt{1-H^2+\frac{2m\Omega}{\rho}(1+H)^2}}, \nonumber \\
\beta = -\left(H-\frac{2m\Omega}{\rho}(1+H) \right) \frac{\alpha^2}{L}, \quad
\gamma = \frac{L}{\alpha}. \end{align}
We recover the Schwarzschild metric in ingoing Eddington-Finkelstein coordinates for $\Omega_{\mathrm{in}}=1$ and $H_{\mathrm{in}}=0$. We use this choice in an inner domain $\rho\leq\rhoi$. To get hyperboloidal scri fixing coordinates outside a compact domain we set on $\rho\geq\rhoo$ (see \cite{Zenginoglu:2007jw})
\be\label{eq:match} \Omega_{\mathrm{out}} = 1-\frac{\rho}{S}, \qquad H_{\mathrm{out}} = 1+\frac{4m\Omega}{S} + 
\frac{(8m^2-C_T^2) \Omega^2}{\rho^2}. \ee
The matching between the domains is performed by a smooth transition function
\be \label{eq:f} f = \left\{
\begin{array}{ll} 0,
 & \rho\leq \rhoi,  \\ 
   f_T, \quad 
  &  \rhoi<\rho<\rhoo,  \\ 
  1, & \rho\geq \rhoo,
\end{array}\right.\ee 
where $f_T$ is \cite{Yunes:2005nn, Vega:2009qb}
\[ f_T = \frac{1}{2}+\frac{1}{2}\tanh \left(\frac{s}{\pi} \left( \tan w(\rho) - \frac{q^2}{\tan w(\rho)} \right)\right), \]
with 
\[ w(\rho):= \frac{\pi}{2} \frac{\rho-\rhoi}{\rhoo-\rhoi}. \]
The free parameter $q$ determines the point $\rho_{1/2}$ at which $f_T$ takes the value $0.5$ and the parameter $s$ determines the slope of $f_T$ at this point.

We set the functions in \eqref{eq:match_3p1} as follows
\[ \Omega = 1- f \frac{\rho}{S}, \quad L = \Omega - \rho \Omega', \quad H = f H_{\mathrm{in}}.\]
The Ricci scalar for the metric \eqref{eq:ss_match} reads
\[ R = \frac{6\Omega}{L^3r^2} (2 L \Omega'(m \Omega+m r \Omega'-r) -
\Omega r(r-2m\Omega)\Omega''). \]
\section{The conformal method}
\label{sec:3}
We take the wave equation with a source term as a toy problem
\be \label{eq:scalar} \widetilde{\Box}_{\tg}\tilde{\Psi}=F(\tilde{x}^\mu,\tg,\tilde{\Psi}).\ee 
The scalar field $\tilde{\Psi}$ is evolved in a background spacetime with metric $\tg$. Here, $\widetilde{\Box}_{\tg}:=\tg^{\mu\nu} \tilde{\nabla}_\mu \tilde{\nabla}_\nu$ where $\tilde{\nabla}$ is the Levi-Civita connection of the metric $\tg$. 

The scalar wave equation is invariant under a change of coordinates but not under a conformal rescaling of the background metric. We need to take the conformal transformation behavior of the scalar wave equation into account to solve it on a conformal background. A conformal rescaling $g=\Omega^2\tg$ of the physical metric $\tg$ with a conformal factor $\Omega>0$ implies the transformation
\[ \left(\Box -\frac{1}{6} R \right) \Psi = \Omega^{-3} \left(\tilde{\Box} -
\frac{1}{6}\tilde{R}\right) \tilde{\Psi}, \quad \mathrm{with} \quad
\Psi = \frac{\tilde{\Psi}}{\Omega}, \] where $R$ and $\tilde{R}$ are
the Ricci scalars of the rescaled and physical metrics $g$ and
$\tg$ respectively. The scalar wave equation (\ref{eq:scalar}) on the background of a solution of the vacuum Einstein equations is equivalent to \be\label{eq:solve} \Box\Psi -\frac{1}{6}
R \Psi = \Omega^{-3} F(x^\mu, \Omega^{-2} g, \Omega \Psi).\ee 
This equation is singular only if the source term does not fall off sufficiently fast towards infinity. This is not a strong restriction because physically motivated problems typically have source terms that have strong fall off or compact support.

The conformal transformation is a minor modification that does not change the principal part of the system. It introduces only an additional source term to the scalar wave equation. The results we present in this paper are expected to apply similarly for any system of equations for test fields with a reasonable conformal transformation behavior. For example, Maxwell and Yang-Mills fields are conformally invariant and therefore, the application of the hyperboloidal method does not require modifications of the geometric equations. The advantage of such a general geometric approach is that, in contrast to the construction of artificial outer boundary conditions, a different numerical technique, background, or equation does not significantly change the application of the method. The situation is rather different for the Einstein equations where the metric plays a double role of background and unknown variable \cite{Friedrich:2002xz}.

\section{Numerical experiments}\label{sec:4}

\subsection{Implementation}\label{subsec:impl}
We solve the conformal wave equation with the Spectral Einstein Code {\texttt SpEC} \cite{SpECWebsite} used for binary black hole evolutions \cite{Scheel:2006gg, Boyle:2007ft}. A pseudospectral technique is applied on each subdomain to evolve the scalar wave equation in time. An important property of the numerical calculation is that it employs spherical boundaries. The numerical outer boundary of the computational domain corresponds to future null infinity, whose cuts are topological spheres, therefore it is advantageous to use spherical grid topology. Our basis functions are Chebyshev polynomials in the radial direction and spherical harmonics in the angular directions. 

Numerical codes that work with multidomain techniques typically employ a global Cartesian system. We give for convenience the transformation of a
spherically symmetric metric of the form (\ref{gen_decomp}) into
Cartesian coordinates. We have coordinates $x_i$ such that
$\rho^2=\delta_{ij} x_i x_j$, where $\delta_{ij}$ is the 3D Kronecker
delta. The 3+1 decomposition of a metric $g$ in Cartesian coordinates
can be written as
\[ g =  \left(-\alpha^2 + \gamma_{ij} \beta^i \beta^j \right)\,d\tau^2 + 2 \gamma_{ij} \beta^i\, d\tau\,dx^j + \gamma_{ij} \,dx^i dx^j.\]
We have
\[ \gamma_{ij} = (\gamma^2-1)\frac{x_i x_j}{\rho^2} + \delta_{ij}, \quad 
\beta^i = \beta \frac{x^i}{\rho}. \] The derivatives are calculated by
\begin{eqnarray*}
\gamma_{ij,k} &=& \left((\gamma^2)_{,\rho}-\frac{2}{\rho}(\gamma^2-1)\right) \frac{x_i x_j x_k}{\rho^3} + \\
&& + \frac{\gamma^2-1}{\rho}\left(\delta_{ik}\frac{x_j}{\rho}+\delta_{jk}\frac{x_i}{\rho}\right), \\
\beta_{i,k} &=& \left(\beta_{,\rho}-\frac{\beta}{\rho}\right)\frac{x_i x_k}{\rho^2} +
\frac{\beta}{\rho}\delta_{ik}, \\ \alpha_{,k} &=& \alpha_{,\rho}
\frac{x_k}{\rho}.
\end{eqnarray*}

We solve a hyperboloidal initial value problem for the conformal wave equation \eqref{eq:solve}  in first order form with auxiliary variables $\Phi_i:= \partial_i \Psi$ and $\Pi:=-1/\alpha(\partial_\tau \Psi-\beta^i \partial_i \Psi)$ as in \cite{Scheel:2003vs, Holst:2004wt}. As initial data we set all fields to zero except $\Pi(0,\rho)$ which is set to an off-centered Gaussian pulse
\be \label{eq:offc}  \Pi(0,\rho) = e^{-|\bf{x}-\bf{x_0}|^2/\sigma^2},  \ee
or to a pure $Y_{lm}$ mode
\be \label{eq:pure_lm} \Pi(0,\rho) = e^{-(\rho-\rho_0)^2/\sigma^2}   Y_{lm}(\vartheta,\varphi). \ee
\subsection{Minkowski spacetime}

\subsubsection{The setup}
The gauge parameters in our coordinatization of the conformal Minkowski spacetime (\ref{eq:con_mink}) are the coordinate location of null infinity $S$, and a parameter $a$ that controls the mean extrinsic curvature of the hyperboloidal surfaces $K$ via $K=3S/a$. To set these parameters suitably we discuss the characteristic structure of Minkowski spacetime in our gauge. The coordinate speeds of in- and outgoing radial light rays are given by
\be\label{eq:chars} c_{\pm} = \pm \frac{\alpha}{\gamma} - \beta. \ee

We get with (\ref{mink_decomp})
\be\label{eq:mink_char} c_+ = \frac{1}{2a}(\rho+S)^2, \qquad c_- = -\frac{1}{2a}(\rho-S)^2.\ee
The outgoing speed near $\scri^+$ is proportional to $S^2/a$. We need to keep the characteristic speeds at the order of unity to avoid strong restrictions on the time step due to the Courant-Friedrichs-Lewy (CFL) condition. We choose the parameters such that $a\sim S^2$ implying $K\sim 1/S$. In our calculations we set $S=20$ and $a=400$. 

The formulae \eqref{eq:mink_char} show that there are no incoming characteristics from the outer boundary at $\rho=S$ implying that no boundary conditions are needed or allowed at the numerical outer boundary. This property of hyperboloidal scri fixing compactification is the main advantage over standard treatments of the outer boundary problem. It simplifies the implementation and makes error controlling and stability considerations trivial. 

\begin{figure}[ht]
\centering 
\includegraphics[width=0.45\textwidth,height=0.23\textheight]{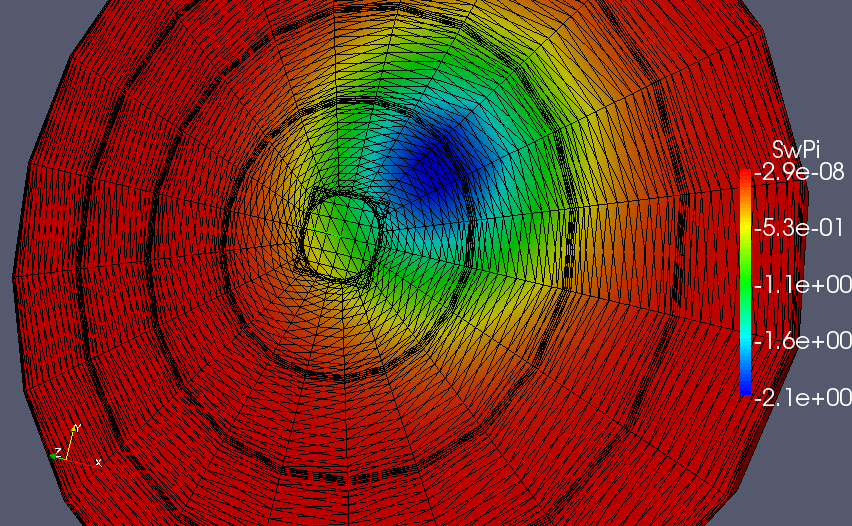}
\caption{The numerical grid topology for calculations in Minkowski spacetime. We have a cube around the origin with the domain $[-2,2]$ in each Cartesian direction and 4 spherical shells extending from $\rho=2$ to future null infinity at $\rho=20$. The colors depict an off-centered Gaussian initial data for $\Pi$ as given in (\ref{eq:offc}). \label{fig:offc}}
\end{figure}

Our simulation domain in the radial direction is given by $\rho\in[0,20]$. The domain structure includes a cube around the origin with the domain $x_i\in[-2,2]$ and 4 spherical shells extending to future null infinity (Fig.~\ref{fig:offc}). 

\subsubsection{Convergence}
\begin{figure}[ht]
 \flushleft
 \includegraphics[width=0.45\textwidth,height=0.27\textheight]{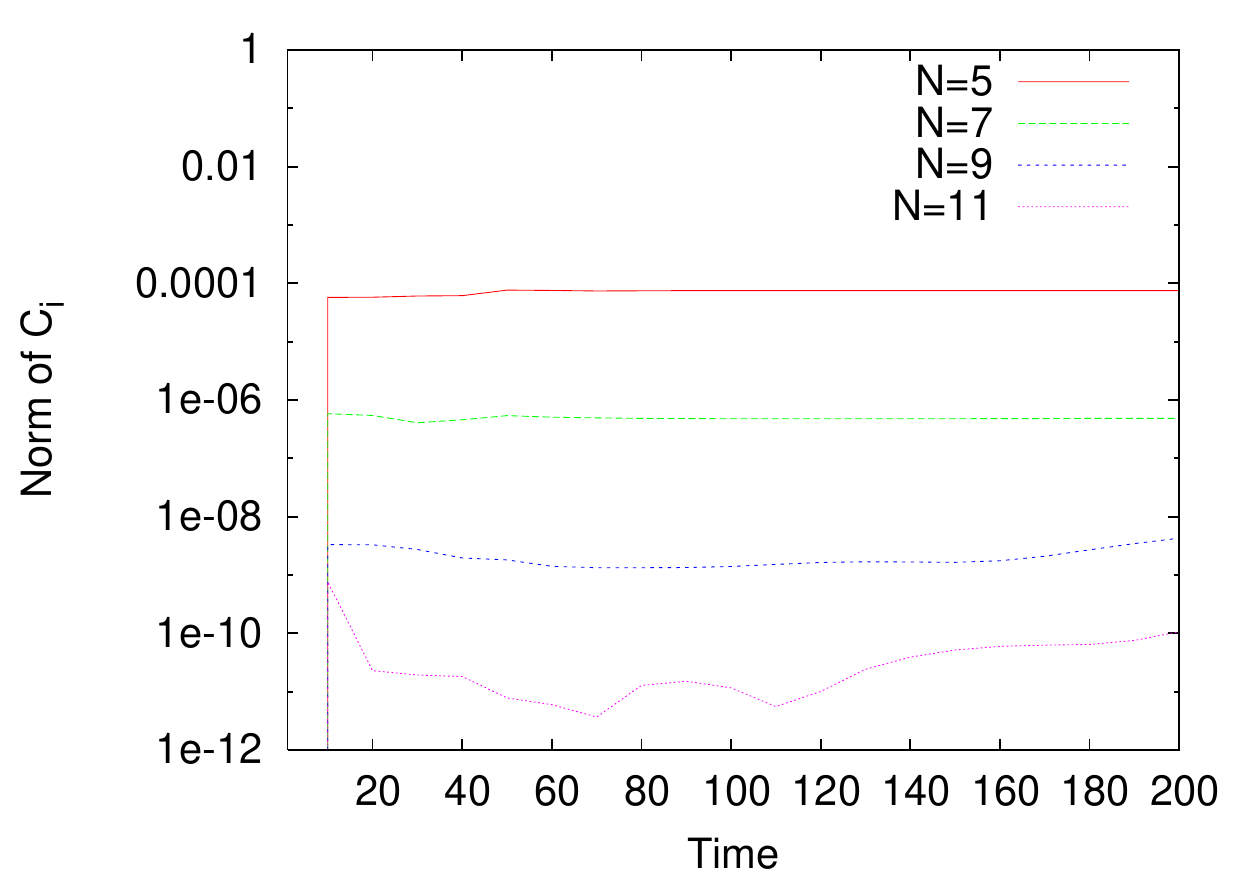}
 \caption{Spectral convergence for the evolution of off-centered initial data in Minkowski spacetime by the $L_2$ norm of the constraint field $C_i$ (\ref{eq:constr}). \label{fig:conv_mink}}
\end{figure}

Spatial truncation errors converge exponentially in a pseudospectral code. We show such spectral convergence in Fig.~\ref{fig:conv_mink} for an evolution with vanishing source and off-centered initial data (Eq.~(\ref{eq:offc}) and Fig.~\ref{fig:offc}). We plot the $L_2$-norm of the constraint field \be \label{eq:constr} C_i = \Phi_i-\partial_i \Psi, \ee in time. The convergence properties of pure $Y_{lm}$-data are similar. The off-centered case provides a stringent test on the accuracy of the code because the evolution is highly asymmetrical and the resolution requirements are higher. 

\subsubsection{Semilinear wave equation}
The source-free wave equation on Minkowski spacetime is a simple test because the wave package leaves the spacetime along the characteristic directions due to the Huygen's principle. When nonlinear source terms are present, however, there is backscatter due to the self interaction of the field and the numerical method needs to deal with nonlinearities. This is a strong test for transparency boundary conditions because they must not eliminate all reflections from the outer boundary but only spurious ones because backscatter near the boundary is part of the solution. 

\begin{figure}[ht]
 \centering 
 \includegraphics[height=0.19\textheight]{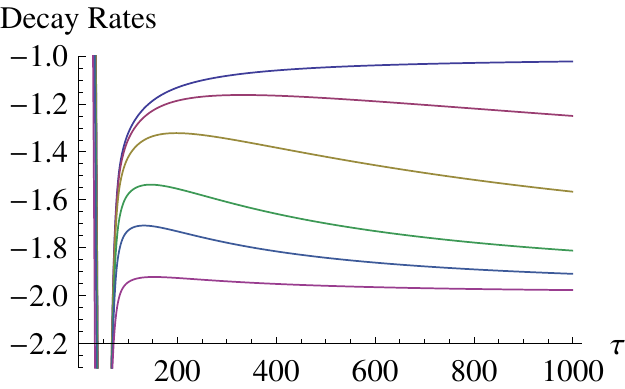}
 \caption{Local decay rates measured by various observers for the cubic wave equation. The locations of the observers vary from null infinity (top curve) to $r=30$ (bottom curve). \label{fig:cubic}}
\end{figure}

We demonstrate the applicability of our method to wave equations with nonlinear source terms by taking a source of the form $F=-\tilde{\Psi}^p$. The minus sign corresponds to focusing \cite{Bizon04}. The conformal wave equation with such a power source term reads
\[ \Box\Psi - \frac{1}{6} R \Psi = -\Omega^{p-3}\, \Psi^p.\]
The right hand side is regular for $p\geq3$. A good measure for the accuracy of the code is whether the backscatter of the field due to self-interaction can be calculated correctly in long time evolutions. We show in Fig.~\ref{fig:cubic} the local decay rates of the field for $p=3$ as measured by various observers. The decay rates at the end of our simulation are $-1.02$ along null infinity and $-1.97$ along $r=30$, very close to the expected asymptotic rates of $-1$ along null infinity and $-2$ along timelike surfaces.

The implementation of nonlinear source terms poses no difficulties and does not change the numerical outer boundary treatment in the hyperboloidal method as long as the terms have sufficiently fast fall off. This behavior should be contrasted to accurate transparency boundary conditions that need to be modified depending on the details of the source terms.

\subsection{Schwarzschild spacetime}
\subsubsection{The setup}\label{sec:ss_pars}
The gauge parameters in our coordinatization of the conformal Schwarzschild spacetime in a CMC foliation (\ref{eq:con_ss}) are the coordinate location of null infinity, $S$, the mean extrinsic curvature of the hyperboloidal surfaces, $K$, and an integration parameter $C$ that influences the causal properties of the CMC foliation. A useful discussion of these parameters can be found in \cite{Buchman:2009ew}.

\begin{figure}[ht]
 \centering 
 \includegraphics[height=0.17\textheight]{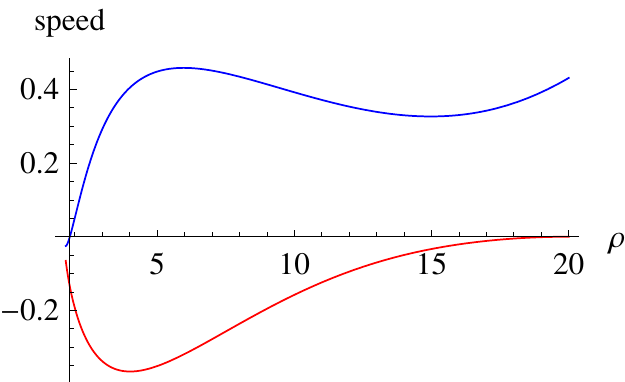}
 \caption{Coordinate speeds of ingoing (red) and outgoing (blue) radial characteristics in Schwarzschild spacetime with respect to a CMC foliation. There are no incoming characteristics to the simulation domain, so no boundary conditions need to be applied. \label{fig:speeds}}
\end{figure}

Evaluation of the characteristic speeds \eqref{eq:chars} at future null infinity shows that the outgoing speed becomes $\frac{2}{9} K^2 S^2$. Therefore we choose the mean extrinsic curvature as $K\sim1/S^2$. We set the coordinate location of future null infinity to $S=20$ as for Minkowski spacetime in the previous section. We further set $m=1,K=0.07$ and $C=1$. The event horizon is located at 
\[ \rho_{\mathcal{H}} = \frac{2 m S}{2m + S}.\]
We use the excision technique by placing the inner boundary inside the event horizon, where all characteristic fields are outgoing and no boundary conditions need to be applied. There is a coordinate singularity inside the event horizon for CMC surfaces on Schwarzschild spacetime given by the zero set of the lapse function \cite{Brill80, Malec:2003dq, Malec:2009hg, Buchman:2009ew}. For the above choice of parameters the singularity is at $\rho_s=1.732$ and the horizon is at $\rho_\mathcal{H}=1.818$. To avoid the singularity we put the numerical inner boundary just inside and very close to the event horizon. Our simulation domain in the radial direction reads $\rho\in[1.8,20]$. The coordinate speeds of in- and outgoing radial characteristics on this domain with the above parameters are plotted in Fig.~\ref{fig:speeds}. The ingoing speed (red curve) vanishes at the outer boundary, whereas the outgoing speed (blue curve)
is non-vanishing and less than unity. The outgoing speed is fairly constant over the grid, which implies even resolution for the outgoing wave across the numerical domain that covers an infinite physical domain. There are no boundary conditions needed because no characteristics enter the simulation domain. The domain structure consists of 9 spherical shells that extend from inside the event horizon at $\rho=1.8$ to future null infinity at $\rho=20$.

\subsubsection{Convergence}
\begin{figure}[ht]
 \flushleft
 \includegraphics[width=0.45\textwidth,height=0.27\textheight]{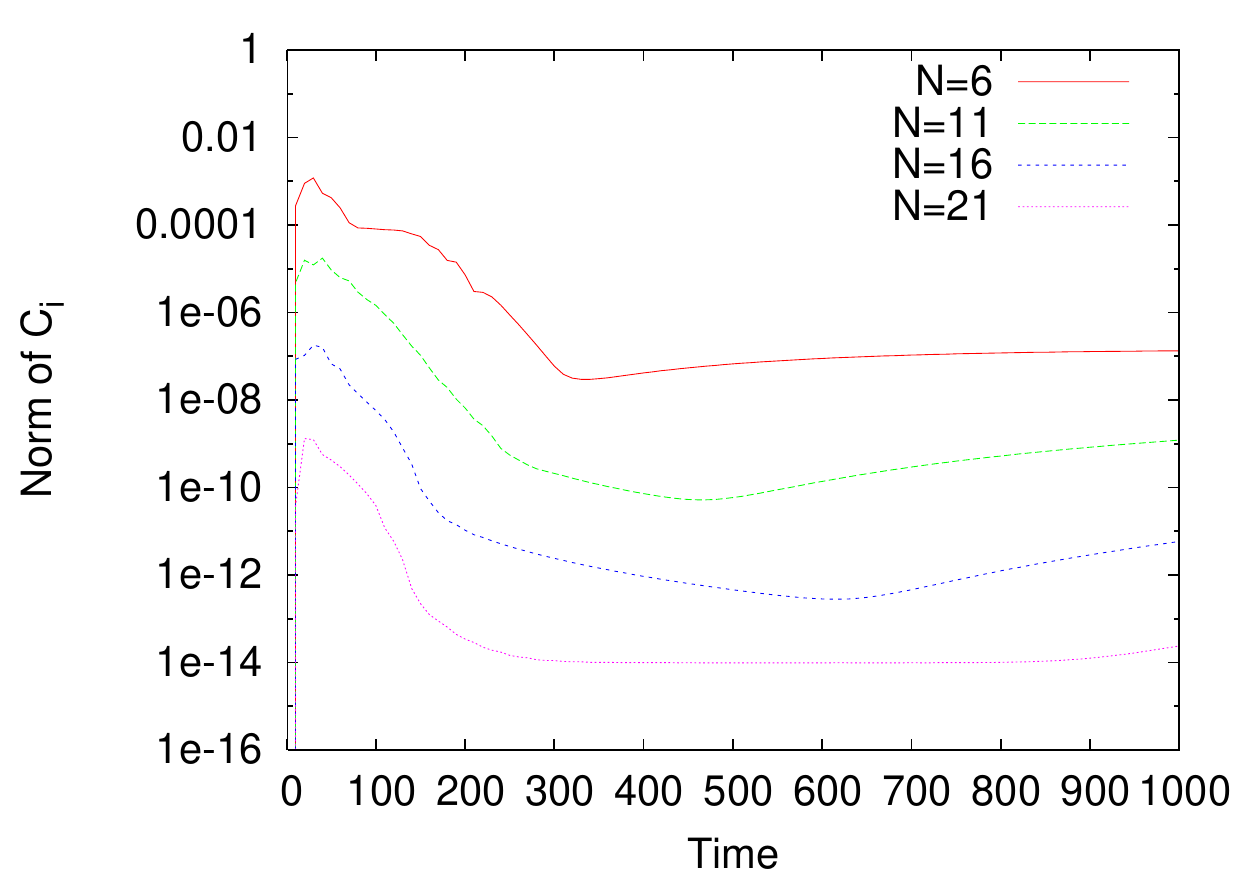}
 \caption{Convergence for the evolution of off-centered initial data in Schwarzschild spacetime by the $L_2$ norm of the constraint field $C_i$ (\ref{eq:constr}). The domain is covered by 9 spherical shells with radial $N$ collocation points each. We fix the angular resolution. \label{fig:conv}}
\end{figure}

We show spectral convergence in Fig.~\ref{fig:conv} for an evolution with vanishing source and off-centered data in Schwarzschild spacetime. For this plot we cover the simulation domain with 9 spherical shells and fix the angular resolution. We perform the radial convergence test with $N$ radial collocation points in each shell. Fig.~\ref{fig:conv} shows the $L_2$ norm of the constraint field $C_i$ defined in  (\ref{eq:constr}) over the whole grid including the outer boundary. 

\subsubsection{Quasinormal mode ringing and tail decay rates}\label{sec:tail}
\begin{figure}[ht]
 \centering 
 \includegraphics[height=0.23\textheight]{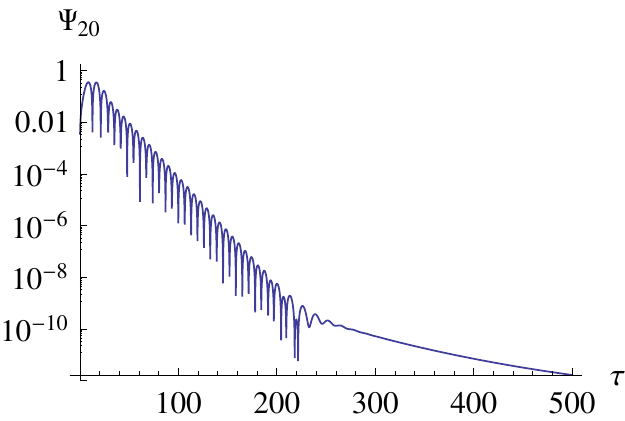}
 \caption{The quasinormal ringing and subsequent polynomial decay in Schwarzschild spacetime for initial data with $l=2,m=0$ in a half-logarithmic plot.\label{fig:qnm}}
\end{figure}

A perturbation of Schwarzschild spacetime evolves in three phases: an initial transient phase that depends on initial data, a quasinormal mode ringing, and a polynomial decay. We show the quasinormal mode ringing and the subsequent polynomial decay in Fig.~\ref{fig:qnm} for the evolution of a pure $l=2,m=0$ initial data in a half-logarithmic plot. The ringing has the form
\be \label{eq:qnm} \Psi(\tau) = a \,e^{-\omega_2 \tau} \sin(\omega_1 \tau + \varphi).\ee Here, $\omega_1$ and $\omega_2$
are the mode frequencies, $a$ is the amplitude and $\varphi$ is
the phase of the wave signal. We measure these parameters by fitting the signal to the above formula using a simple least squares method on the interval $\tau\in[60m,120m]$. We find $\omega_1 = 0.48389$ and $\omega_2=0.09661$. These numerical values are very close to those obtained by using Leaver's continued fraction method \cite{Leaver85, Leaver:1986gd}, which read $\omega_1 = 0.48364$ and $\omega_2=0.09676$. The fitting error in the mode frequencies dominate the numerical discretization error of the solution.

In general, it is difficult to calculate the polynomial decay rates accurately with a 3D code. Furthermore, numerical calculations of such decay rates usually measure the rates near the black hole, which are different from the rates in the far-field zone \cite{Gundlach94a}. The correct and accurate numerical calculation of asymptotic decay rates as measured by idealized observers can only be achieved in compactified schemes that include null infinity \cite{Gundlach94a, Purrer:2004nq, Zenginoglu:2008wc}. 

\begin{figure}[ht]
 \centering 
 \includegraphics[height=0.19\textheight]{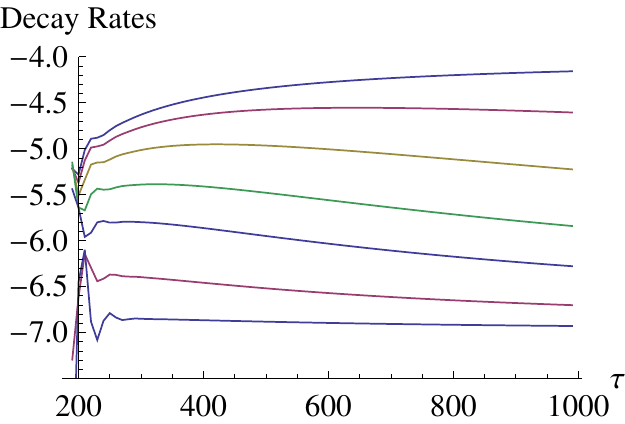}
 \caption{Local decay rates measured by various observers for the evolution of pure $l=2,m=2$ initial data. The locations of the observers vary from null infinity (top curve) to $r=30m$ (bottom curve). \label{fig:l2}}
\end{figure}

We show in Fig.~\ref{fig:l2} the local power indices as measured by observers ranging from near the black hole to future null infinity for initial data with a pure $l=2,m=2$ mode, which is expected to be the dominant mode for gravitational waves. This mode decays asymptotically in time with rates of $-4$ along null infinity and $-7$ along timelike surfaces. Fig.~\ref{fig:l2} indicates that these asymptotic decay rates can be reproduced cleanly in a 3D long time evolution with the hyperboloidal method. The coordinate location of the numerical outer boundary is 20. In the standard approach this would imply that Fig.~\ref{fig:l2} shows the decay rates for 50 crossing times without any contamination from the outer boundary. Of course, the notion of a crossing time does not apply in our case because the incoming propagation speed from the outer boundary vanishes.

\subsubsection{Matching}
The matching is a rather arbitrary way of gluing hypersurfaces together; therefore there are many free parameters to be set. We fix the coordinate location of null infinity to $S=20$ as in the previous computations. To set the constant $C_T$ in \eqref{eq:match} we consider the outgoing characteristic speed at null infinity, which reads $S^2/C_T^2$. So we set $C_T=S$.

\begin{figure}[ht]
 \flushleft
 \includegraphics[width=0.45\textwidth,height=0.27\textheight]{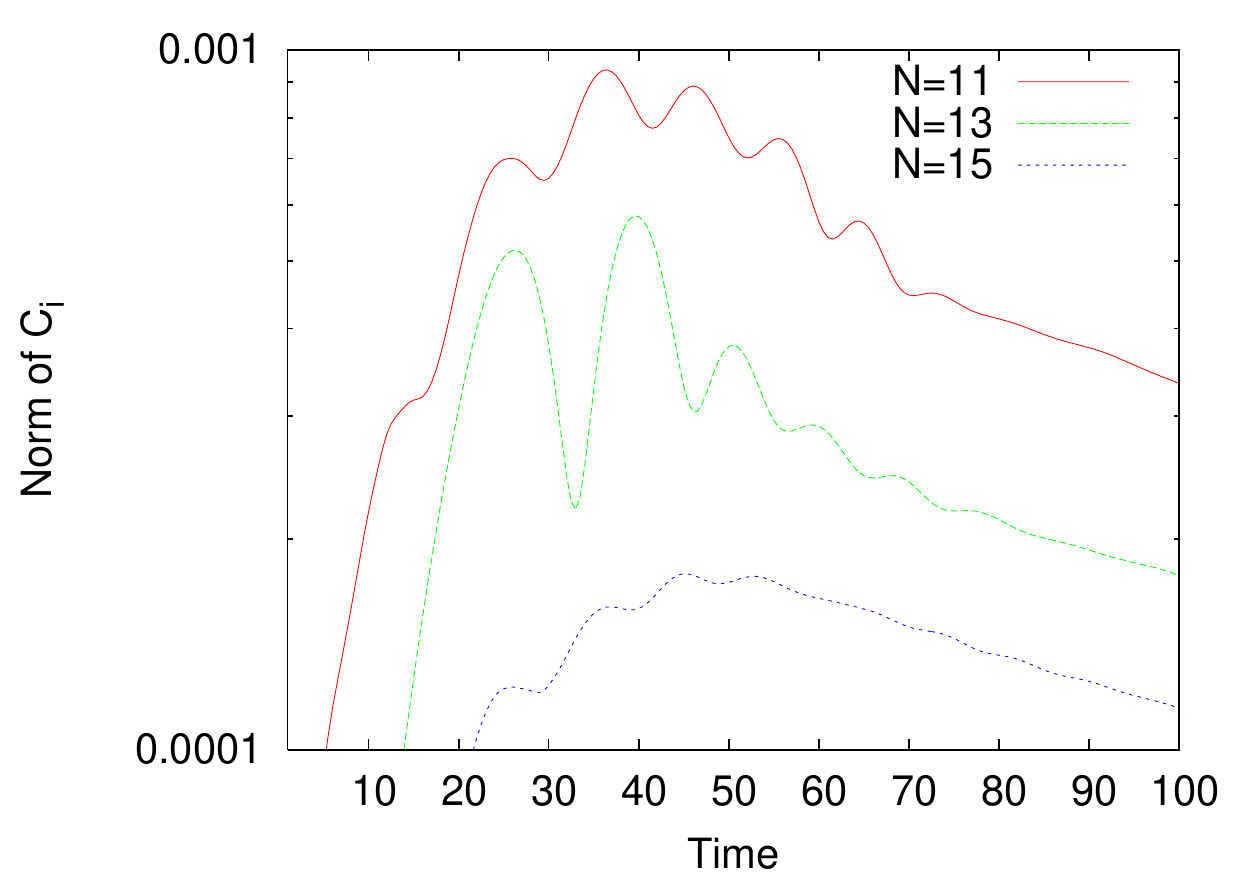}
 \caption{Convergence for the evolution of a pure $l=2,m=2$ initial data in Schwarzschild spacetime with a matched foliation. \label{fig:conv_match}}
\end{figure}

The formula for the height function derivative \eqref{eq:match}  cannot be used everywhere because it violates the spacelike condition \eqref{eq:sp_cond}. With $m=1$ and the above choice of parameters, the condition is violated at $r=12.5$. The inner matching radius must be larger than this value. We can move this point closer to the black hole in terms of the grid coordinate by performing the compactification earlier than the bending up of the height function. To achieve this we introduce parameters $\rhoi^\Omega,\rhoo^\Omega,\rhoi^H,\rhoo^H$ so that the matching of the physical coordinate to the compactifying coordinate is performed independently than the matching of the ingoing Eddington-Finkelstein time surfaces to asymptotically hyperboloidal time surfaces. A good choice of parameters that we used for the evolutions underlying the convergence plot in Fig.~\ref{fig:conv_match} is $\rhoi^\Omega=4,\rhoo^\Omega=10, \rho^H=10, \rho^H=15$.

Fig.~\ref{fig:conv_match} shows the convergence properties of matched evolutions for initial data with a pure $l=2,m=2$ mode. On the one hand, it is reassuring that matching can be performed stably in combination with pseudospectral methods. On the other hand, Fig.~\ref{fig:conv_match} shows that the errors are about four orders of magnitude bigger than the errors in Fig.~\ref{fig:conv}, so the matching is very inefficient compared to CMC calculations. This difference can be explained as follows. The spatial redshift effect \cite{Zenginoglu:2007jw, Zenginoglu:2008uc, Zenginoglu:2009ey} leads to highly efficient numerical calculations in combination with high order discretization schemes such as pseudospectral methods as we demonstrated in the previous sections. With matching, this effect is active only in part of the domain. Instead, the additional structure from matching needs to be resolved and is a source of numerical error. One can expect that lower order or finite element type numerical methods will work better with matching.

The form of the metric that we implement for matching is formally singular at future null infinity where the conformal factor $\Omega$ vanishes \eqref{eq:match_3p1}. Instead of applying the analytic limits at the outer boundary, we employ in the outermost shell Gauss-Radau collocation points, which do not involve the point at the outer boundary. This facilitates a natural extrapolation within the pseudospectral method and allows us to evolve the wave equation without numerically evaluating the formally singular coefficients. A similar formal singularity is present in the conformal Einstein equations. A strategy to implement the hyperboloidal approach in combination with the generalized harmonic formulation of the conformal Einstein equations requires the numerical evaluation of such formally singular terms at the outer boundary \cite{Zenginoglu:2008pw}. Our results suggest that a similar extrapolation method for the conformal Einstein equations may be successful. We emphasize, however, that the model problem studied here is simpler due to the decoupling of the formal singularity from the evolution equations. A successful implementation of conformal Einstein equations that does not require the evaluation of such terms has been presented by Rinne using a constrained scheme in axisymmetry \cite{Rinne:2009}.

\section{Conclusions}
The hyperboloidal method allows us to numerically calculate solutions to hyperbolic PDEs on unbounded domains thereby avoiding the outer boundary problem and cleanly solving the radiation extraction problem. The approach in this paper is based on the conformal method \cite{Penrose65}, the hyperboloidal initial value problem \cite{Friedrich83a} and the hyperboloidal scri fixing compactification of asymptotically flat spacetimes \cite{Zenginoglu:2007jw} in combination with the Spectral Einstein Code {\texttt SpEC} developed by the Caltech-Cornell collaboration \cite{SpECWebsite}. Our results obtained for the toy model of a scalar field can be expected to hold equally for other test fields such as Maxwell or Yang-Mills fields. Numerical tests with off centered and pure spherical harmonic data and nonlinear source terms in Minkowski and Schwarzschild spacetimes support our conclusions.

The efficiency of the hyperboloidal method in combination with pseudospectral methods is striking. All computations for this paper have been performed on a laptop without parallelization. Decay rates as strong as $-7$ can be accurately calculated in 3D with less than 200 radial collocation points to cover an infinite domain of the Schwarzschild spacetime. 

There are many advantages of the hyperboloidal method beyond efficiency, especially in comparison to standard outer boundary treatments. First of all, the question of well-posedness and stability of boundary conditions becomes trivial because there are no boundary conditions. In standard approaches, boundary data needs to be chosen to ensure the transparency of the boundary and to model the exterior domain. This calculation is typically complicated and falls out completely in the hyperboloidal method. The geometric nature of the method makes it largely independent of the details of numerical schemes, whereas it is a nontrivial task to implement certain accurate boundary conditions for high order numerical discretizations. The hyperboloidal method requires minor modifications for a large class of equations and background spacetimes. Achieving good accuracy for the boundary treatment of each of the examples we presented took decades of work and further improvements are exceedingly complicated, whereas the hyperboloidal method applies easily to problems with a suitable asymptotic behavior. The method is also practical. There is no overhead in software implementation and therefore no need to consider runtime cost at the boundary in comparison to the interior grid work. There is no need for error controlling because there are no errors other than numerical discretization. In addition, we have access to the radiation signal as measured by idealized observers at future null infinity.

The time transformation can be regarded as a disadvantage of the hyperboloidal method because certain applications require a specific coordinate system in a compact domain. The matching technique avoids the time transformation in such a compact domain. We observe, however, that matching is less efficient than a CMC foliation. This is in accordance with expectations because with matching there is additional structure that needs to be resolved by the numerical scheme. It also introduces many free parameters complicating the application of the method. In cases where the use of a specific coordinate system is necessary, the decision whether to employ the matching technique will depend on the importance of boundary conditions, time scale of the simulation, and even the numerical scheme. Further research in this area may show that matching works well with certain numerical schemes, but for now our experiments suggest that matching should be avoided if possible.

Currently, matching is required for the application of the hyperboloidal method in Kerr spacetime. A study of CMC foliations of Kerr spacetime would be beneficial for improving the accuracy and efficiency of numerical computations on a rotating black hole background. The analogous problem for nonrotating black holes has been solved in 1980 by Brill, Cavallo, and Isenberg \cite{Brill80}. It seems that time has come to attack the problem for rotating black holes.

A further disadvantage of the hyperboloidal method is the requirement of a spherical grid topology at the outer boundary. Cuts of future null infinity naturally have spherical topology. Previous studies of hyperboloidal compactification with Cartesian grids report spurious reflections of outgoing signals through null infinity \cite{vanMeter:2006mv}. Therefore this technical requirement may be a limiting factor for applying the hyperboloidal method in numerical codes employing Cartesian grids.

Eventually, the main problem that we would like to tackle is the hyperboloidal initial value problem in general relativity. Our results should be useful both for regular conformal field equations \cite{Friedrich81b, Husa01, Frauendiener04} and for conformal Einstein equations \cite{Zenginoglu:2008pw, Moncrief:2008ie, Rinne:2009}. Specifically relevant for the full problem in general relativity are the following observations: free evolution is very accurate and efficient so that even strong decay rates can be resolved in 3D without using large computational resources; the spatial redshift effect of hyperboloidal foliations works well with pseudospectral methods; nonlinear source terms can be handled without difficulties; the coordinatization of a compact domain can be chosen freely but matching is an additional source of numerical error; and the numerical evaluation of formally singular terms at the outer boundary using Gauss-Radau collocation points leads to stable evolutions. These results may play a fundamental role in improving the accuracy and efficiency of numerical relativistic calculations with the hyperboloidal approach.
\vspace{-0.37cm}
\begin{acknowledgments}
\vspace{-0.25cm}
AZ gratefully acknowledges Eitan Tadmor and support by
the Center for Scientific Computation and Mathematical Modeling
(CSCAMM).  We thank Luisa Buchman, Enrique Pazos, Oliver Rinne, Saul
Teukolsky, and Manuel Tiglio for discussions.  This work was supported
by the NSF grant 07-07949 at CSCAMM, by the Marie Curie Transfer of Knowledge contract MTKD-CT-2006-042360 at Krak\'ow, and by grants from the Sherman Fairchild
Foundation, NSF Grants No. PHY-0652952 and No. PHY-0652929 and NASA
Grant No. NNX09AF96G at Cornell.
\end{acknowledgments}

\end{document}